\documentclass{pap}
\usepackage{graphicx,latexsym,isolatin1}
\usepackage{psfrag,amsmath}
\usepackage{mathrsfs}
\usepackage[mathcal]{euscript}
\usepackage[english]{babel}

\newcommand{\media}[1]{\left\langle  #1 \right\rangle}
\newcommand{\p}[1]{\left({#1}\right)}
\newcommand{\pq}[1]{\left[{#1}\right]}
\newcommand{\pg}[1]{\left\{{#1}\right\}}

\newcommand{\D}[2]{\frac{\partial #1}{\partial #2}}

\title{Work probability distribution in  single molecule experiments}
\shorttitle{Work probability distribution}
\author{Alberto Imparato
\thanks{Corresponding author. E-mail: imparato@na.infn.it}
\and Luca Peliti
\thanks{Associato INFN, Sezione di Napoli.}}
\institute{Dipartimento di Scienze Fisiche and Unit\`a INFM, Universit\`a ``Federico II'', Complesso Universitario di Monte S. Angelo, I-80126 Napoli (Italy).}
\pacs{87.14.Gg}{DNA, RNA}
\pacs{82.37.-j}{Single molecule kinetics}
\pacs{82.37.Rs}{Single molecule manipulation of proteins and other biological molecules}
\begin{document}
\maketitle
\begin{abstract}
We derive and solve a differential equation satisfied by the probability distribution of the work done on a single biomolecule in a mechanical unzipping experiment.
The unzipping is described as a thermally activated escape process in an energy landscape. The Jarzynski equality is recovered as an identity, independent of the pulling protocol. This approach allows one to evaluate easily, by numerical integration, the work distribution, once a few parameters of the energy landscape are known.
\end{abstract}
\section{Introduction}
The introduction of micromanipulation techniques has dramatically
improved our knowledge of physical and chemical properties of
biological molecules. Such techniques have been used to probe the
structure of proteins \cite{rgo,png,ash,nst,pbp} and nucleic acids
\cite{bus1}. A typical experiment consists of pulling the
free end of a biomolecule  with a controlled force, while its
end-to-end distance is measured at the same time. It has been
suggested that the study of the kinetics of bond breakage under
different loading rates can provide many informations on the
molecule internal structure, and in particular  allows one to
measure the strength of the molecular bonds, and to associate to
them a position along the molecular structure \cite{ev1}. The
loading-rate dependent kinetics experiments on biomolecules  have
been interpreted in terms of thermally activated escape from bond
states over a succession of energy barriers,  along a
one-dimensional energy landscape \cite{ev2,denis,Str,albepjb}.

Usually, because of technical limitations, the molecule  pulling process is characterized by
 time scales much faster than the typical molecular relaxation time. This prevents the possibility to perform the experiment in quasiequilibrium conditions and thus  to obtain direct measurements of the thermodynamic state variables.
This difficulty can be overcome by exploiting the remarkable equality derived by Jarzynski \cite{jar,jar1} (Jarzynski equality, JE), and extended by Crooks \cite{cro}, which allows one to obtain the free energy difference $\Delta F$ between two equilibrium states by evaluating the average of $\exp(-\beta W)$, where $W$ is the work performed on the system during the thermodynamic transformation:
\begin{equation}
e^{-\beta \Delta F}=\media{e^{-\beta W}}\, .
\label{je}
\end{equation}
Here $\beta=1/k_B T$, and the average on the rhs is performed over all possible realizations of the process starting from the equilibrium ensemble at a given temperature $T$.
Since this equality holds in general, the information of $\Delta F$ can be gathered also by processes so fast that they do not leave the system at equilibrium.
On the other hand, the process must be sampled a large number of times in order to obtain a reliable estimate of $\Delta F$, as discussed in \cite{bus3,rit1}.
 In the same references it is argued that the number of pulling experiments needed to achieve  good statistics in the estimate of $\Delta F$ increases as the transformation are made more irreversible, for example as the pulling rate is increased.
Furthermore the JE gives no information on the probability distribution of work: while the JE magnifies the
rare trajectories with $W<\Delta F$, a direct measurements of the  probability distribution for these value
of $W$ is rather difficult. On the other hand, the effect of such trajectories on the  average thermodynamic variables might become significant when the energies involved range between a few and tens of $k_B T$'s, as in the case of molecular bonds.

The aim of this paper is thus to provide an effective method to evaluate the work probability distribution
of a molecular pulling process, which can be described as an escape process in an energy landscape, once the main features of the landscape are known. Such a method can be applied independently of the
irreversibility of the process, i.e., of the pulling protocol, and for any value of the maximum force.

The article is organized as follows. We first describe our model of escape process.
We then introduce a set of  differential equations describing the time evolution
of the work distribution probability $\phi(W,t)$ and show that the JE follows as an identity.
We next consider a simple case of escape process, and discuss the behaviour of the
work distribution probability $\phi(W,t)$ obtained for it.

\section{The model}\label{sec1}
\begin{figure}[h]
\center
\input{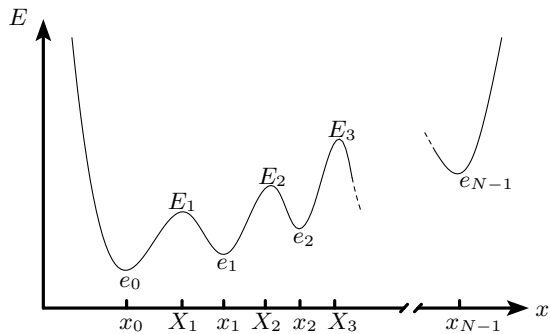}
\caption{Generic energy landscape, representing the succession of bonds in a biomolecule.}
\label{land}
\end{figure}

In a typical unzipping experiment, a force $f(t)$ is applied on one
end of the biomolecule, and its elongation $x(t)$ is monitored. Within
some limits, $x(t)$ can be considered as a collective coordinate for
the system, spanning a one-dimensional (free) energy landscape
$E(x)$. This energy landscape will in general be characterized
by a set of $N$ minima of energy $e_i$ at position $x_i$,
 with $i=0, \dots, N-1$, and by a set of $N-1$ maxima of energy $E_j$
at position $X_j$ with $j=1,\dots, N-1$, see figure \ref{land}.
The escape from one minimum $e_i$ over the next maximum $E_{i+1}$ can be viewed
as the breaking of a given molecular bond.
The evolution of the collective coordinate will be governed by a
stochastic process satisfying detailed balance. If however the
variations of the  energy $E$ between maxima and minima are large
enough with respect to $k_\mathrm{B}T$, it is possible to describe this
stochastic process as a Markov process with discrete states,
corresponding to the extremals of the  energy, while the
transition probabilities will be given by the Kramers expression.
For simplicity, we shall now consider only transitions between nearest neighbour energy minima  $j\rightarrow i$, i.e. $i=j\pm 1$. However, we will argue that the results we obtain in the present paper are rather general and can be extended to the case of transitions between any pair of states $j$ and $i$.
Thus the transition rate from one minimum $j$ to a neighboring one $i$,
over the corresponding energy barrier, is given by
\begin{equation}
 k_{i j}(t)=  \omega_0 \exp\pg{-\beta \pq{E_{ij} -e_j-f(t)(X_{ij}-x_j)}}\, ,
\end{equation}
where the energy $E_{ij}$ and the position $X_{ij}$ of the barriers are respectively given by
\begin{equation}
E_{i j}=E_{\max\pg{i,j}}\, ,
\qquad
X_{i j}=X_{\max\pg{i,j}}\, ,
\end{equation}
and where $\omega_0$ is some attempt rate whose value depends on the system characteristics.
Let $p_i(t)$ be the probability that the system is in state  $i$ at time $t$:
using a Kramers formalism the time evolution of these probabilities is described
by the following set of differential equations
\begin{equation}
\D{p_i(t)}{t}=\sum_j\pq{ k_{ij}(t)p_j(t)-k_{ji}(t)p_i(t)}\, .
\label{pt}
\end{equation}
The jump process, described by the set of equations (\ref{pt}), is clearly  Markovian
and  it preserves the starting equilibrium ensemble for any fixed value of $f$.
These conditions, together with the assumption that the energies involved in the process are finite,
are sufficient for the JE to be recovered for the dynamics described by eqs.~(\ref{pt}), as shown by Jarzynski \cite{jar, jar1} and Crooks \cite{cro}.

In the following $Z_t$ will denote the
 the partition function of the canonical ensemble at temperature $T$ and with an applied force equal to $f(t)$:
\begin{equation}
Z_t=\sum_i\exp\pg{-\beta\pq{e_i-f(t)x_i}}\, .
\end{equation}

\section{Work probability distribution}\label{sec2}
The probabilities $p_i(t)$ introduced in the previous section, are
not sufficient to describe the probability distribution of the
work done on the system. We are thus interested in the joint
probability distribution $\phi_i(W,t)$ that the system is in state
$i$ at time $t$, while the total work done on it is equal to $W$.

If the system is in the state $i$ at time $t$,
the work done $\delta W$ in the time interval $\delta t$ is given by \protect{\footnote{In the present paper we use the ensemble where the force is the externally controlled parameter, while the extension of the molecule $x_i$ fluctuates. This implies that the effective work done on the system, as the force increases of $\delta f$, is $\delta W=-x_i \delta f$. In pulling experiments where an optical tweezer is  used  to pull one of the free ends of the molecule, the actual controlled parameter is the position of the focus of the optical trap.
In this situation both the applied force and the molecule extension fluctuate.
Here we assume that the fluctuations in $f(t)$ are small, even if the conditions under which  this assumption is valid deserve further investigation.}
\begin{equation}
\delta W=-\delta t \D{f(t)}{t}x_i \, .
\label{dW}
\end{equation}
By expanding to first order in $\delta t$ and $\delta W$, the probability distribution $\phi_i(W+\delta W, t+\delta t)$,
it is easy to verify that
this function satisfies the set of differential Chapman-Kolmogorov  equations:
\begin{equation}
\D{\phi_i(W,t)}{t}=\sum_j\pq{ k_{ij}(t)\phi_j(W,t)-k_{ji}(t)\phi_i(W,t)}+\dot f(t)x_i \D{\phi_i(W,t)}{W}\, .
\label{phit}
\end{equation}
The probability $p_i(t)$ and the  probability distribution $\phi_i(W,t)$ are connected by the relation
\begin{equation}
\int d W \phi_i(W,t)=p_i(t)\;,
\end{equation}
provided that $\phi_i(W,t)$ satisfies the additional boundary conditions
$\phi_i(\pm \infty,t)=0\, .$
By integrating both sides of equation (\ref{phit}) with respect to $W$,
one recovers the set of differential equations (\ref{pt}).

We want now to show that the JE is satisfied for a system characterized by a generic
energy landscape as that represented in fig.~\ref{land},
whose work probability distributions evolve according to eq.~(\ref{phit}).
Let $\phi^l_j(W,t)$ denote the probability distribution of the work in the state $j$ at time $t$,
when the system initial state is $l$. They satisfy the set of equations (\ref{phit})
with the initial conditions
\begin{equation}
\phi^l_j(W,t=0)=\left\{ \begin{array}{ll} \delta (W), & \textrm{if}\quad  j= l\, ; \\
                                      0, & \textrm{if}\quad  j\neq l\, ;
            \end{array}
         \right.
\label{initialc}
\end{equation}
Thus the average of $e^{-\beta W}$ up to time $\tau$ is given by
\begin{equation}
\media{e^{-\beta W}}_\tau=\sum_l \frac{\exp\p{-\beta e_l}}{Z_0}\int e^{-\beta W} \pq{\sum_j \phi^l_j(W,\tau)} d W=\int e^{-\beta W} \phi(W,\tau) d W\, ,
\label{jarw}
\end{equation}
where
\begin{equation}
\phi(W,t)=\sum_{lj} \frac{\exp\p{-\beta e_l}}{Z_0} \phi^l_j(W,t)\, ,
\end{equation}
is the total work distribution probability.
Our goal is to show that the rhs of eq.~(\ref{jarw}) is equal to  $\exp(-\beta \Delta F)$.
 Let us define the quantity $M_i(t)$ as
\begin{equation}
M_i(t)=\int e^{-\beta W} \sum_l \frac{e^{-\beta e_l}}{Z_0} \phi^l_i(W,t) d W \, ,
\label{mi}
\end{equation}
the rhs of eq.~(\ref{jarw}) can thus be written as
\begin{equation}
\int e^{-\beta W} \phi(W,t)d W=\sum_i M_i(t).
\label{phimi}
\end{equation}
By taking the time derivative of $M_i(t)$, and substituting eqs.~(\ref{phit}) and (\ref{mi}), we obtain
\begin{eqnarray}
\frac{d M_i(t)}{d t}&=&
\int e^{-\beta W} \sum_l \frac{e^{-\beta e_l}}{Z_0}  \frac{\partial}{\partial t}\phi^l_i(W,t) d W\label{primaW} \\
&=& \sum_j \pq{k_{ij}(t)M_j(t)-k_{ji}(t)M_i(t)}+\int e^{-\beta W} \sum_l \frac{e^{-\beta e_l}}{Z_0}
\dot f(t)x_i \partial_W\phi_i^l(W,t) d W\, .
\label{terzaM}
\end{eqnarray}
The last term in eq.~(\ref{terzaM}) can be integrated by parts, and using
the boundary condition that $\phi^l_i(W,t)$ goes to zero faster than $\exp(\beta W)$
as $W\rightarrow -\infty$ (which follows from the existence of the average in eq.~(\ref{je})),
we find  that $M_i(t)$ satisfies
\begin{equation}
\frac{d M_i(t)}{d t}= \sum_j \pq{k_{ij}(t)M_j(t)-k_{ji}(t)M_i(t)}+\beta \dot f(t)x_i M_i(t)\, ,
\label{dmi}
\end{equation}
with the initial conditions
$M_i(0)=e^{-\beta e_i}/Z_0$,
which follow from the initial value conditions for the functions $\phi^l_j(W,t)$, eq.~(\ref{initialc}), and from the definition of $M_i(t)$, eq.~(\ref{mi}).
It is easy to verify that the functions
\begin{equation}
M_i(t)=\frac{\exp\pq{-\beta (e_i-f(t)x_i)}}{Z_0}
\end{equation}
are the solutions of the set of equations (\ref{dmi}) with the corresponding initial conditions.
Substituting this last result into eq.~(\ref{phimi}), we finally find
\begin{equation}
\int e^{-\beta W}\phi(W,\tau)d W=\sum_i \frac{\exp\pq{-\beta (e_i-f(\tau)x_i)}}{Z_0}=\frac{Z_\tau}{Z_0}\, ,
\label{lastw}
\end{equation}
which verifies the JE.
\section{Work distribution probability: a case study} \label{sec3}
We consider now a simple system with three minima and two maxima in the energy landscape. The values of the energy minima and maxima, expressed in $k_BT$ units,  and their positions, expressed in nm,  are $\{e_0=0, \, x_0=0\}, \, \{E_1=10, \, X_1=0.6\}, \,\{ e_1=6, \, x_1=0.8\}, \,\{ E_2=16, \, X_2=1.8\},\, \{ e_2=12, \, x_2=2\}$.

We consider here the case of pulling processes where the force increases linearly with time, i.e. $f=r\cdot t$, where $r$ is the pulling rate.
In order to simplify the notation, let
\begin{equation}
\tilde \phi^l_j(W,t)=\frac{e^{-\beta e_l}}{Z_0}\phi^l_j(W,t)\, .
\end{equation}
The work probability distribution $\phi(W,t)$ is thus given by
\begin{equation}
\phi(W,t)=\sum_{l,j} \tilde \phi^l_j(W,t)\, .
\label{phitot}
\end{equation}
We obtain the probability $\phi(W,t)$, for two values of the pulling rate $r=1, \, 10$ pN/s, by solving the set of equations (\ref{phit}) with the parameter choice as indicated above, taking the initial value conditions as given by eq.~(\ref{initialc}) and the attempt frequency $\omega_0=4.4\,  \mathrm{s}^{-1}$.
With this choice of parameters,
we obtain the zero-force transition rate $k_0=\omega_0 \exp\pq{-\beta (E_i -e_{i-1})}=2\times 10^{-4}\,  \mathrm{s}^{-1}$, in agreement with the  zero-force transition rate found in the mechanical unfolding of a simple RNA molecule \cite{bus1}.

The results at different times are shown in figure \ref{conf_r1}
for $r=1$ pN/s and $r=10$ pN/s.
In order to analyze the behavior of the tails of the work distribution,
the quantity $\phi(W,t)$ for $r=1$ pN/s is plotted in a log-linear form in figure  \ref{tail_r1}, for the smallest and the largest times considered.
Recalling that $\phi(W,t)$ is the sum of the work probability distributions $\tilde \phi^l_j(W,t)$ along all the possible trajectories, see eq.~(\ref{phitot}),
in the same figures the main contributions to the total  distribution probability of the work are plotted.
Inspection of figure  \ref{tail_r1}
suggests that at small times and for small values of $|W|$  the work probability distribution
is dominated by those trajectories which start and finish in the state $x_0$.
On the other hand, at any time, at large $|W|$,  $\phi(W,t)$ is dominated
by those trajectories which finish in the rightmost state $x_2$, as expected.
In particular, in the very large $|W|$ regime, the distribution $\tilde \phi^2_2(W,t)$
determines the behavior of  $\phi(W,t)$.
\begin{figure}[h]
\center
\psfrag{p}[bc][bc][.7]{$\phi(W,t)$}
\psfrag{W}[cl][cl][.7]{$W$}
\psfrag{p1}[bl][bl][1.]{$\phi(W,t)$}
\psfrag{W1}[cl][cl][1.]{$W$}
\psfrag{t1}[br][br][.8]{$t=3.3$ s}
\psfrag{t2}[br][br][.8]{$t=33$ s}
\psfrag{t3}[br][br][.8]{$t=50$ s}
\psfrag{t4}[br][br][.8]{$t=63$ s}
\includegraphics[width=7cm]{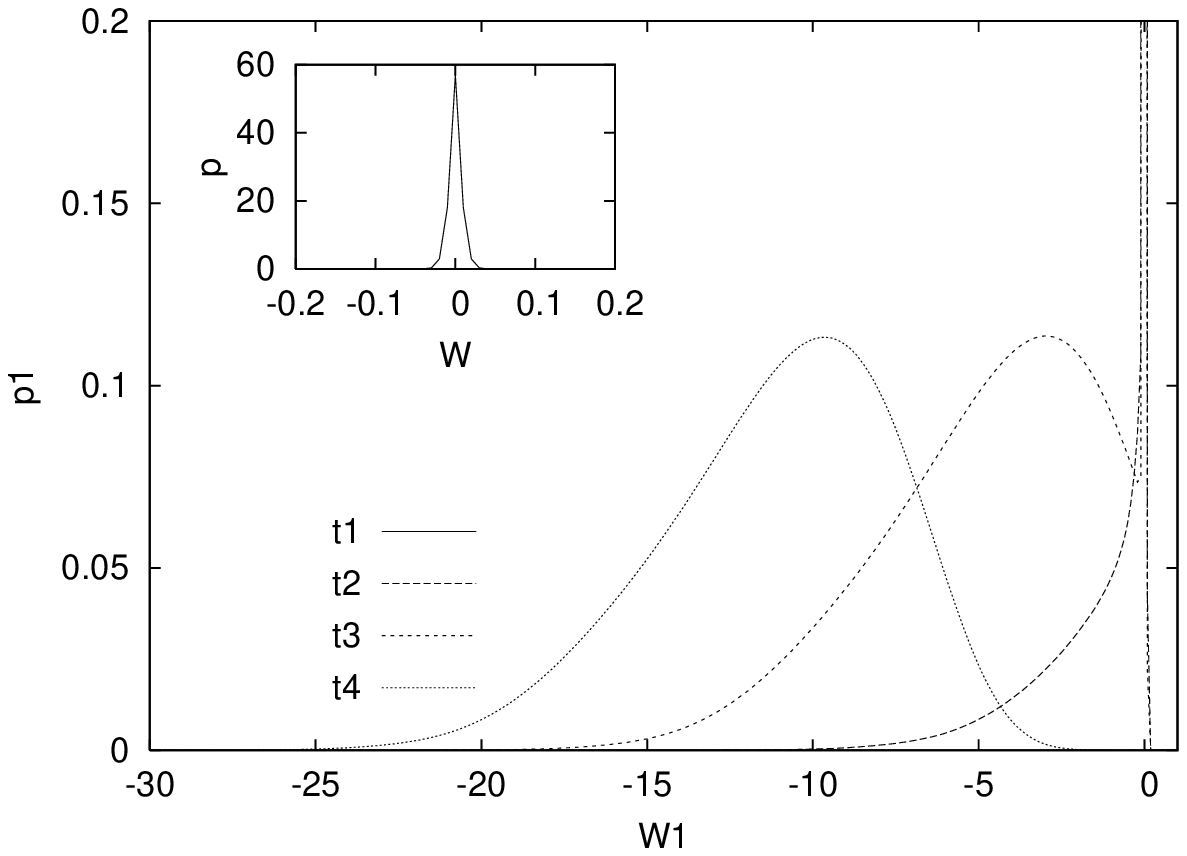}
\psfrag{p}[bc][bc][.7]{$\phi(W,t)$}
\psfrag{W}[cl][cl][.7]{$W$}
\psfrag{p1}[bl][bl][1.]{$\phi(W,t)$}
\psfrag{W1}[cl][cl][1.]{$W$}
\psfrag{t1}[br][br][.8]{$t=.33$ s}
\psfrag{t2}[br][br][.8]{$t=3.3$ s}
\psfrag{t3}[br][br][.8]{$t=5.0$ s}
\psfrag{t4}[br][br][.8]{$t=6.3$ s}
\includegraphics[width=7cm]{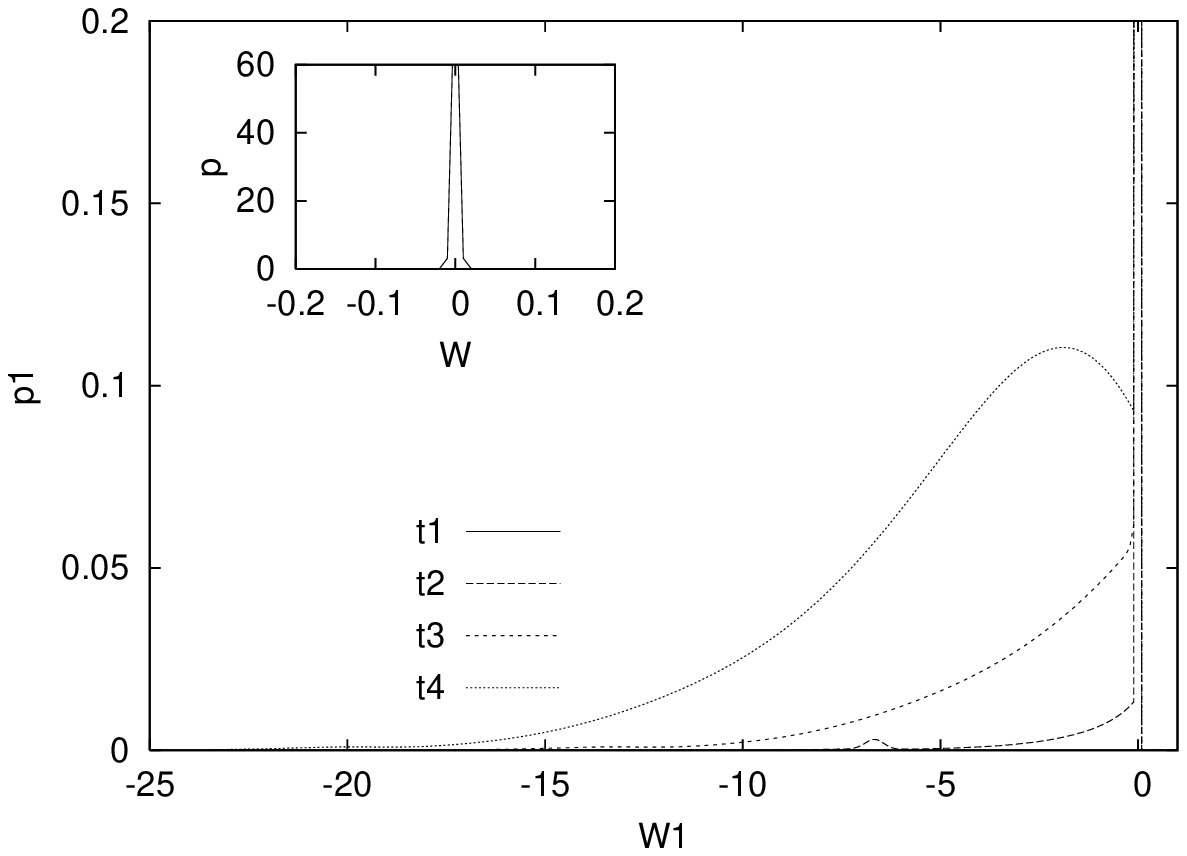}
\caption{Plot of the work distribution probability $\phi(W,t)$, as defined by eq.~(\ref{phitot})
as a function of the work $W$ at different times. Left: $r=1~\mathrm{pN}/\mathrm{s}$. Right:
$r=10~\mathrm{pN}/\mathrm{s}$. Inset: $\phi(W,t)$ at the smallest time here shown (left: $t=3.3$ s; right: $t=.33$ s).}
\label{conf_r1}
\end{figure}

\begin{figure}[h]
\center
\psfrag{W}[cl][cl][1.]{$W$}
\psfrag{a}[br][br][.9]{$\phi(W,t_1)$}
\psfrag{b}[br][br][.9]{$\tilde\phi^0_0(W,t_1)$}
\psfrag{c}[br][br][.9]{$\tilde\phi^1_1(W,t_1)$}
\psfrag{d}[br][br][.9]{$\tilde\phi^2_2(W,t_1)$}
\includegraphics[width=7cm]{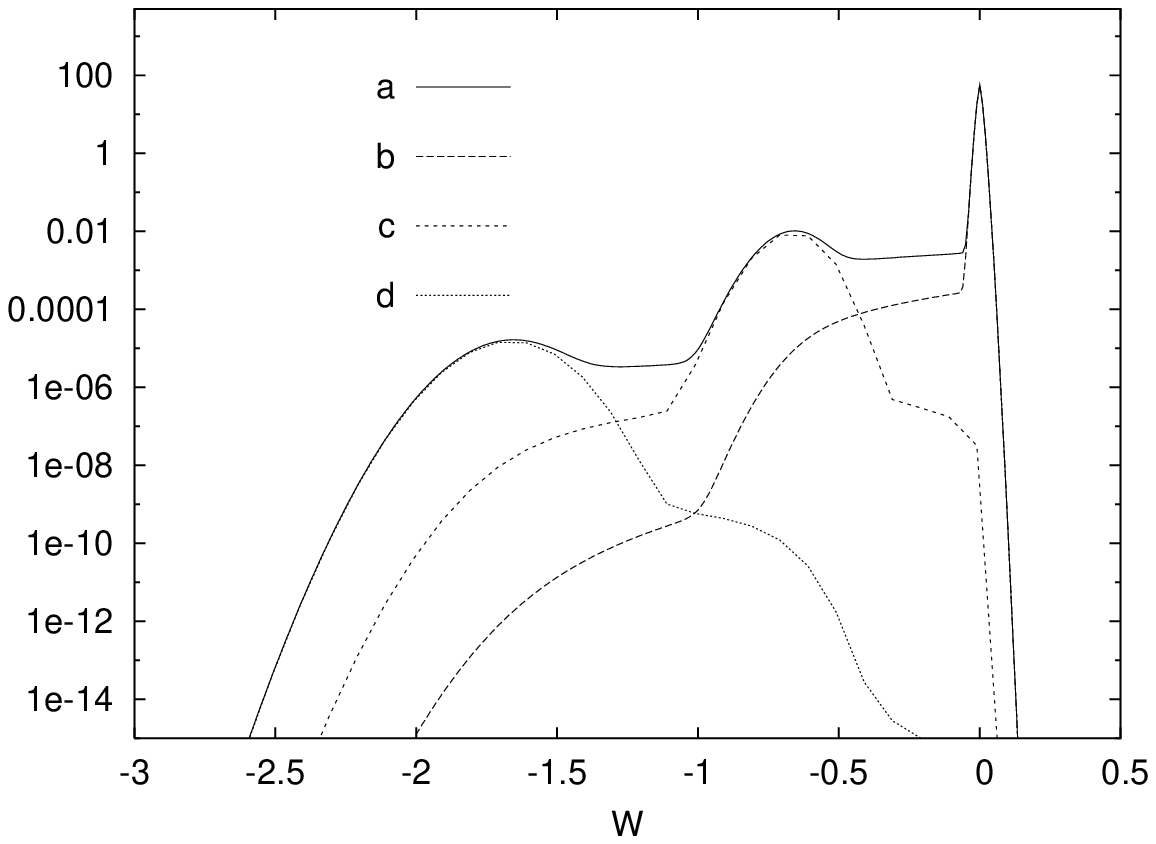}
\psfrag{a}[br][br][.9]{$\phi(W,t_2)$}
\psfrag{b}[br][br][.9]{$\tilde\phi^0_0(W,t_2)$}
\psfrag{c}[br][br][.9]{$\tilde\phi^1_1(W,t_2)$}
\psfrag{d}[br][br][.9]{$\tilde\phi^2_2(W,t_2)$}
\psfrag{e}[br][br][.9]{$\tilde\phi^0_2(W,t_2)$}
\includegraphics[width=7cm]{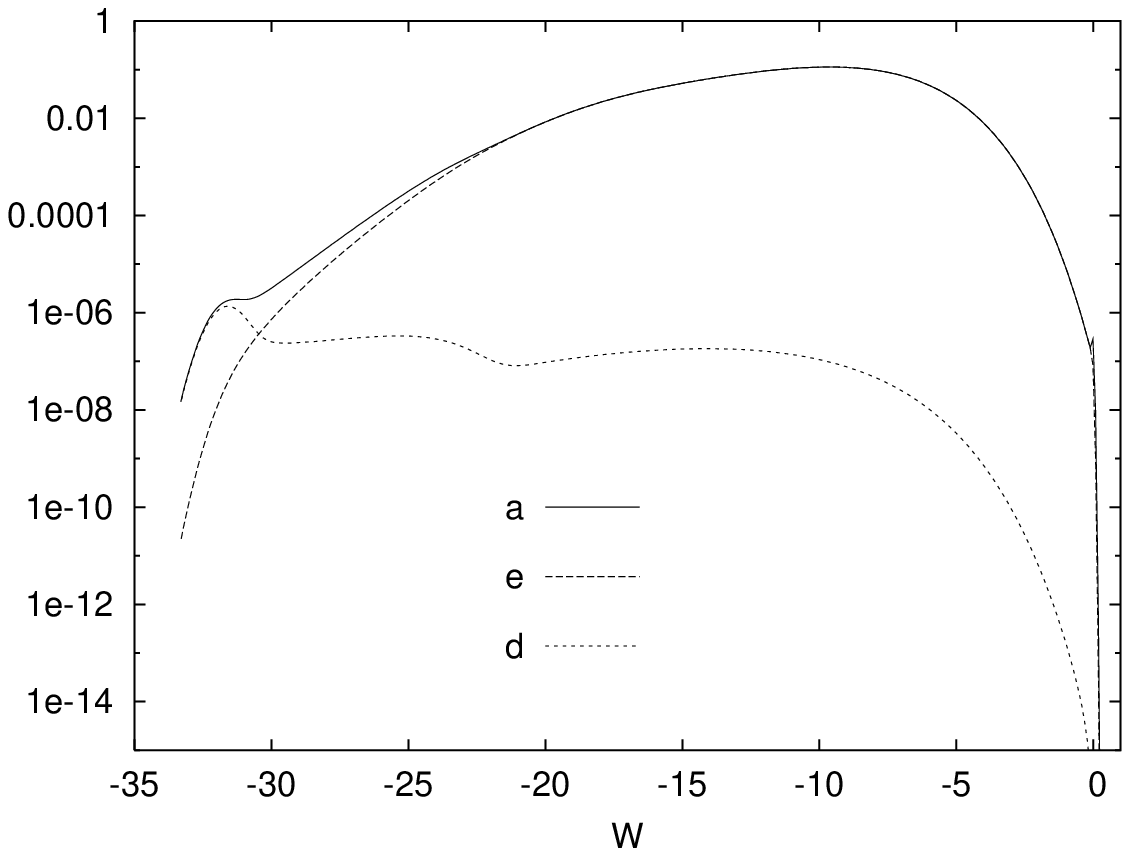}
\caption{Log-linear plot of the work distribution probability
$\phi(W,t)$ as a function of $W$ for $r=1~\mathrm{pN}/\mathrm{s}$, and of the main contributions to the sum
on the right hand side of eq.~(\ref{phitot}), at short times ($t_1=3.3$) and at long times ($t_2=63$ s).}
\label{tail_r1}
\end{figure}

Thus, at any time, a given distribution function $\tilde \phi_j^l(W,t)$ dominates the behavior of the
total work distribution probability $\phi(W,t)$ in  a given range of $W$.
In each of such work ranges, the function $\phi(W,t)$ can be well fitted by a gaussian distribution (fits not shown), i.e., the  function $\phi(W,t)$ is a superposition
of several gaussian functions, each with different mean and variance, rather than
a single gaussian function.
The gaussian distribution of the work, expected for a slowly perturbed system as
discussed in \cite{rit1,rit2}, is thus not recovered here  for the pulling
rate $r=1$ pN/s, which is a lower limit in the pulling experiments of biomolecules.
\section{Discussion}
In the present paper we have introduced a theory which allows us to obtain the work probability distribution
performed on a biomolecules during a pulling experiment.
The pulling process has been described as a jump process over a  succession of energy barriers.
This approach allow us to describe the time evolution of the  population of the minimal energy states using a set of stochastic differential equations. By discretizing such process, we show that the probability
of the discrete trajectories satisfies the Jarzynski equality independently of the pulling protocol.
The derivation of the JE within the current formalism can be viewed as the discrete-state-space analogue of the analysis in Sec.~II of ref. \cite{jar1} and in Sec.~IV of ref. \cite{jar3}.

We then derive the set of differential equations describing the time evolution of the work probability distributions for each trajectory connecting two arbitrary initial and final states of the system.
By summing up these distribution functions over all the initial and final states, we obtain  the total work probability distribution $\phi(W,t)$.
We show that the JE is recovered for such a distribution function: in this sense, the average over an infinite number of trajectories of eq.~(\ref{je}), is replaced by an average over the probability distribution $\phi(W,t)$ as in eq.~(\ref{jarw}).
In deriving this results, we have assumed that the main features of the energy landscape, i.e., the height and the positions of the wells and of the barriers, are known. For  real biomolecules these quantities have to be measured independently for the function $\phi(W,t)$ to be numerically calculated.
By considering a very simple energy landscape, we calculate numerically the work probability distribution, and find that it exhibits a non-gaussian behaviour, being rather a superposition of gaussian functions, each corresponding to distinct trajectories between different initial and final state.

We  point out that our results  still hold if we consider a jump process
where transitions between any pair of states are allowed.
In fact, as long as the transition rates $k_{ij}$ satisfy the detailed balance condition,
eqs.~(\ref{primaW},\ref{lastw}) are still verified,
and thus one recovers the results obtained
for the  case of jumps between successive states.

\acknowledgments
We are grateful to F. Ritort for introducing us
to the topic and for interesting discussions.

\end{document}